\documentclass[reprint,superscriptaddress,amsmath,amssymb,aps,prb]{revtex4-1}
\usepackage{graphicx}
\usepackage{float}% Include figure files
\usepackage{dcolumn}% Align table columns on decimal point
\usepackage{bm}% bold math
\usepackage{pifont}
\usepackage{hyperref}% add hypertext capabilities
\usepackage[mathlines]{lineno}

\usepackage{soul}
\usepackage{color}

\begin{document}

\title{Manipulating Ferrimagnets by Fields and Currents}

\author{Mingda Guo}
\thanks{These two authors contributed equally to this work.}
\affiliation{Department of Physics and Astronomy, University of California, Riverside, California 92521, USA}
\author{Hantao Zhang}
\thanks{These two authors contributed equally to this work.}
\affiliation{Department of Electrical and Computer Engineering, University of California, Riverside, California 92521, USA}
\author{Ran Cheng}
\email[Correspondence should be sent to: ]{rancheng@ucr.edu}
\affiliation{Department of Electrical and Computer Engineering, University of California, Riverside, California 92521, USA}
\affiliation{Department of Physics and Astronomy, University of California, Riverside, California 92521, USA}

\begin{abstract}
Ferrimagnets (FIMs) can function as high-frequency antiferromagnets while being easy to detect as ferromagnets, offering unique opportunities for ultrafast device applications. While the physical behavior of FIMs near the compensation point has been widely studied, there lacks a generic understanding of FIMs where the ratio of sublattice spins can vary freely between the ferromagnetic and antiferromagnetic limits. Here we investigate the physical properties of a two-sublattice FIM manipulated by static magnetic fields and current-induced torques. By continuously varying the ratio of sublattice spins, we clarify how the dynamical chiral modes in a FIM are intrinsically connected to their ferro- and antiferromagnetic counterparts, which reveals unique features not visible near the compensation point. In particular, we find that current-induced torques can trigger spontaneous oscillation of the terahertz exchange mode. Compared with its realization in antiferromagnets, a spin-torque oscillator using FIMs not only has a reduced threshold current density but also can be self-stabilized, obviating the need for dynamic feedback.
\end{abstract}

\maketitle

\section{Introduction}
Antiferromagnetic (AFM) spintronics has been a surging frontier interconnecting fundamental physics and electrical engineering, bringing about novel functionalities for next-generation magnetic devices \cite{BaltzReview2018,Jungwirth2018,GomonayReview,Wang2017}. As a defining advantage, AFM materials can typically be operated at the terahertz (THz) regime that outpaces the established ferromagnetic (FM) materials by more than two orders of magnitude in speed, which holds great promise for ultrafast device applications not even theoretically possible in established paradigms. However, one intractable problem hindering the progress of AFM spintronics is that the vanishing magnetization, which should otherwise be a merit, makes the detection of AFM order, hence the reading mechanism in device engineering, very difficult.

One way out of this dilemma is to use ferrimagnets (FIMs) in which magnetic moments are anti-aligned similar to their AFM counterparts, while maintaining an uncompensated magnetization very easy to detect. In FIMs, spin dynamics breaks up into different branches well separated in frequency. For example, a two-sublattice FIM admits two dynamical modes of opposite chirality: a right-handed mode lying in the GHz frequency range and a left-handed mode lying in the THz frequency range~\cite{SeKwon2020,Kim2021,gurevich1996magnetization}. When the low-frequency spin dynamics is excited, a FIM behaves just as an ordinary FM material as if the sublattice spins are locked together. For a long time, archetypal FIMs such as YIG have been treated this way~\cite{brataas2020spin,yang2018fmr}. Therefore, to achieve the functionalities of AFM materials using FIMs, the key is to leverage the high-frequency spin dynamics known as the \textit{exchange mode} (or optical mode), in which the noncollinearity among sublattice spins becomes prominent such that the spin dynamics is driven by the strong exchange interaction.

In existing studies, various AFM-like behavior of FIMs has been recognized in the vicinity of the compensation points~\cite{finley2020spintronics,Stanciu2006,Okuno2020,Kim2017,Kim2017PRB,IvanovReview2019,SeKwon2020,Wangsness1955}, where either the spins or the magnetic moments of different sublattices cancel within a magnetic unit cell (i.e. when the system approaches the AFM limit). However, there lacks a general understanding of FIMs that allows us to freely vary the ratio of sublattice spins between the FM and the AFM limit. Such a generic picture can not only reveal the unique features not visible near the compensation point but also unifies the spin dynamics in FIMs with that in FM and AFM materials.

In this paper, we study the static and dynamical properties of a representative two-sublattice FIM manipulated by magnetic fields and current-induced torques. By varying the ratio of sublattice spins continuously, we clarify the intrinsic connection between the chiral modes in FIM and their FM and AFM counterparts. Depending on the direction of a driving current, the low-frequency FM mode and the high-frequency exchange mode can be selectively excited. While the former evolves into a magnetization switching, the latter evolves into a steady-state oscillation in the THz regime. We find that the threshold current density triggering the auto-oscillation of the exchange mode depends on the ratio of sublattice spins, where a minimum is identified between the FM and the AFM limits. In addition to the reduced threshold, THz spin-torque oscillators realized in FIMs can be self-stabilized, i.e., jumping directly into the spin-flop configuration beyond the threshold is avoided, whereas their AFM counterparts call for a dynamic feedback mechanism to achieve this goal.

The paper is organized as follows. In Sec.~\ref{sec:static}, we study the static properties of our model FIM by minimizing the free energy, where the equilibrium spin configuration is obtained as a function of the magnetic field. In Sec.~\ref{sec:modes}, we solve the uniform chiral modes under a varying magnetic field, where we identify an important chirality flip below the spin-flop transition. In Sec.~\ref{sec:STO}, we study the current-induced dynamics of the model FIM including its spontaneous oscillations, where we plot a series of dynamical phase diagrams benchmarking against the FM and the AFM limits. These findings are concluded in Sec.~\ref{sec:discuss} with physical remarks and outlooks.

%--------------------------------------------------------------------------------------------------------------------------------------------------%

\section{Static Properties}\label{sec:static}

Let us consider a two-sublattice FIM characterized by two macrospin variables $\bm{S}_1$ and $\bm{S}_2$ that are antiferromagnetically coupled. Correspondingly, the sublattice magnetic moments are $\bm{M}_1=\gamma_1\bm{S}_1$ and $\bm{M}_2=\gamma_2\bm{S}_2$, where the gyromagnetic ratios $\gamma_1$ and $\gamma_2$ may in general be different and the competition between these unequal magnetizations switches at the temperature of angular momentum compensation~\cite{coey_2010,KittelReview,ScottReview}. However, in typical FIMs consisting of Fe, Co, and Gd ions, $\gamma_1$ and $\gamma_2$ differ by only a few percent. Therefore, to a good approximation, we only consider $\gamma_1=\gamma_2=\gamma$, which serves as an acceptable simplification to capture the most essential behavior of FIMs. Correspondingly, the ratio of sublattice spins and that of sublattice magnetization can be described by the same parameter $\xi=|S_2/S_1|=|M_2/M_1|$. Throughout this paper, we assume that $S=|S_1|+|S_2|$ is a constant, which follows that $M_s\equiv\gamma S=|M_1|+|M_2|$ is a constant. Accordingly,
\begin{subequations}
\begin{align}
  \frac{|M_1|}{M_s} = \frac{1}{\xi +1},\\
  \frac{|M_2|}{M_s} = \frac{\xi}{\xi +1}.
\end{align}
\label{eq:spinratio}%
\end{subequations}
By definition, $\xi$ can be varied continuously from $0$ to $1$. When $\xi\rightarrow0$, one sublattice vanishes and the system effectively becomes a FM material. On the other hand, if $\xi\rightarrow1$, the two sublattices fully compensate and thus the system becomes an AFM material. The wide range of $\xi$ covers not only a broad class of FIMs with different chemical compositions but also physically interesting regions not accessible by varying temperature, revealing profound implications not necessarily achievable in real materials.

\begin{figure}
    \centering
    \includegraphics[width = \linewidth]{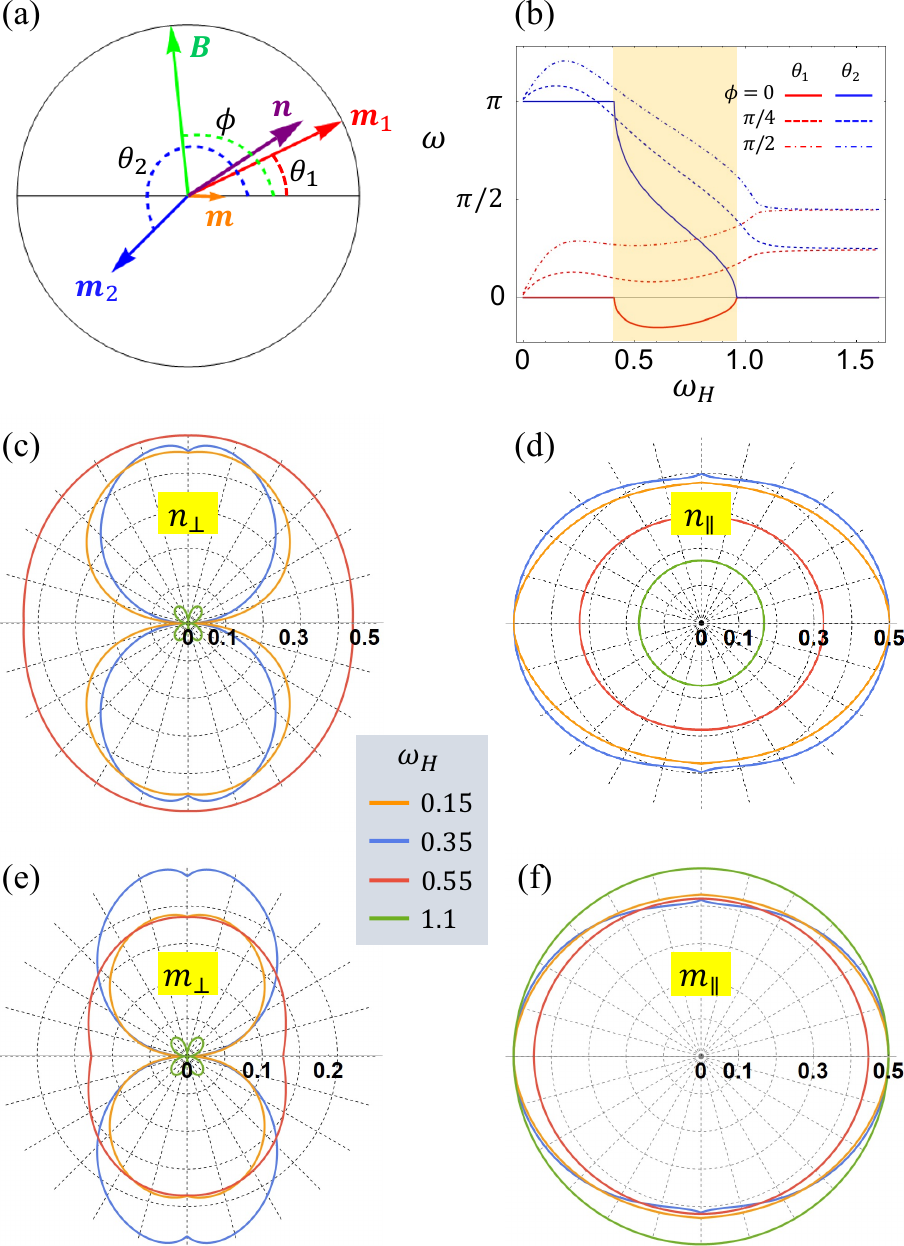}
    \caption{(a) Schematic illustration of system geometry. The vectors $\bm{m}_1$, $\bm{m}_2$, and $\bm{B}$ are characterized by $\theta_1$, $\theta_2$, and $\phi$ relative to the positive $\hat{x}$ direction. The unit vector of total magnetization $\bm{m}=(\bm{m}_1+\bm{m}_2)/2$ and the Néel vector $\bm{n}=(\bm{m}_1-\bm{m}_2)/2$ are represented by orange and purple arrows.
    (b) $\theta_1$ (red) and $\theta_2$ (blue) as functions of the field strength $\omega_H$ along different field angle $\phi$, where a SF phase (shaded) is clearly seen for $\phi=0$.
    (c)-(f) The perpendicular and parallel components of $\bm{m}$ and $\bm{n}$ with respect to the field direction when $\phi$ varies from $0$ to $2\pi$ at different field strengths. Orange and blue curves are below the SF transition, red curves are within the SF phase, and green curves are in the spin-flip phase. $\xi=0.5$ and $\omega_A=0.08$ are used for all plots.}
    \label{fig:angles}
\end{figure}

We study the ground state in terms of two dimensionless vectors $\bm{m}_1=\bm{M}_1/M_s$ and $\bm{m}_2=\bm{M}_2/M_s$. (Note: they are \textit{not} unit vectors) To this end, we consider the free energy density
\begin{align} \label{eq:free_energy}
&\mathcal{E} = J \bm{m}_1 \cdot \bm{m}_2 - \frac{A}2 [(\bm{m}_1 \cdot \hat{x})^2 + (\bm{m}_2 \cdot \hat{x})^2] \notag\\
        & + \frac{K}2 [(\bm{m}_1 \cdot \hat{z})^2 + (\bm{m}_2 \cdot \hat{z})^2] - M_s\bm{H}_0 \cdot (\bm{m}_1+\bm{m}_2),
\end{align}
where $\bm{H}_0$ is the external magnetic field and $J$, $A$, and $K$ are the AFM exchange coupling, the easy-axis and the hard-axis anisotropy, respectively, all taken to be positive and having absorbed the factor of $S$.  A sublattice-specific form of anisotropy is hard to derive and not universal in FIMs, so here we follow the common simplification that treats the anisotropy energies for both sublattices on an equal footing~$\cite{Kim2021,SeKwon2020,lisenkov2019subterahertz}$. In addition, since typically $K\gg A$, inducing an out-of-plane rotation requires an unreasonably large magnetic field. So, to study the ground states, we restrict our discussion to an in-plane field parametrized by an azimuthal angle $\phi$ as illustrated in Fig.~\ref{fig:angles}(a). When $\phi$ varies, $\bm{m}_1$ and $\bm{m}_2$ experience in-plane rotations characterized by $\theta_1$ and $\theta_2$. Therefore, we are able to simplify $\mathcal{E}$ as a function of three angles:
\begin{align}
    \mathcal{E} = &\frac{\hbar \xi}{(\xi+1)^2}\left[ \omega_J\cos(\theta_1-\theta_2)-\frac{\omega_A}{2}(\cos^2\theta_1+\xi^2\cos^2\theta_2)\right] \notag
   \\
    &- \frac{\hbar\omega_H}{\xi+1} [\cos(\phi-\theta_1) + \xi \cos(\phi- \theta_2)], \label{Ham}
\end{align}
where $\hbar\omega_J=J$, $\hbar\omega_A=A$, $\hbar\omega_H=H_0M_s$, and a constant term proportional to $K$ has been omitted.  The equilibrium state can then be obtained by minimizing $\mathcal{E}$ with respect to $\theta_1$ and $\theta_2$ under given field strength $\omega_H$ and field angle $\phi$. Here, we scale all parameters into angular frequencies because this is a convenient convention for the simulation of spin dynamics to be discussed later. We further choose $\omega_J=1$ such that $\omega_A$ and $\omega_H$ are both normalized to the exchange energy.

Before changing $\xi$, we first look into the system's reaction to an increasing magnetic field and how it differs from the AFM case for a particular value $\xi=0.5$ (i.e., $S_2=S_1/2$). Figure~\ref{fig:angles}(b) plots $\theta_1$ and $\theta_2$ with a sweeping field strength $\omega_H$ at different angles. When $\phi=0$, i.e., $\bm{H}_0$ is along the easy axis, the two spins remain unchanged until a spin-flop (SF) transition takes place at around $\omega_H=0.4$, where they undergo an abrupt rotation towards a canted configuration. Beyond this SF point, the two spins fold towards the field direction until they are fully polarized at about $\omega_H=0.95$, after which the system enters the spin-flip phase ~\cite{Clark1968,Rad1986,Beuerle1994,Becker2017} . This behavior is quite similar to that in AFM systems. The difference is that $\theta_1$ and $\theta_2$, in spite of the sudden rotation at the SF point, are continuous everywhere, while in the AFM limit they both jump by nearly $\pi/2$ across the SF point. For finite field angles, the SF phase boundaries are smeared out in a way that not only $\theta_1$ and $\theta_2$ but also their derivatives with respect to $\omega_H$ become continuous.

Next, we study the ground state configuration by changing the field angle $\phi$ continuously from $0$ to $2\pi$. Figures~\ref{fig:angles}(c)-\ref{fig:angles}(f) plot the absolute values of parallel and perpendicular components of $\bm{m}=(\bm{m}_1+\bm{m}_2)/2$ and $\bm{n}=(\bm{m}_1-\bm{m}_2)/2$ relative to the field direction as functions of $\phi$ for four different field strengths (two below the SF transition, one within the SF phase, and one in the spin-flip phase). While the parallel components $\bm{m}_{\parallel}$ and $\bm{n}_{\parallel}$ do not exhibit significant variations with $\phi$, the perpendicular components $\bm{m}_{\perp}$ and $\bm{n}_{\perp}$ show distinct features in different phases: there are two lobes in the collinear phase, a single lobe in the SF phase, and four tiny lobes in the spin-flip phase. Regardless of the field strength, all components repeat themselves in $\phi\in[0,\pi]$ and $\phi\in[\pi,2\pi]$, so in the following we will focus on the range $\phi\in[0,\pi]$ only.

\begin{figure}[t]
    \centering
    \includegraphics[width = \linewidth]{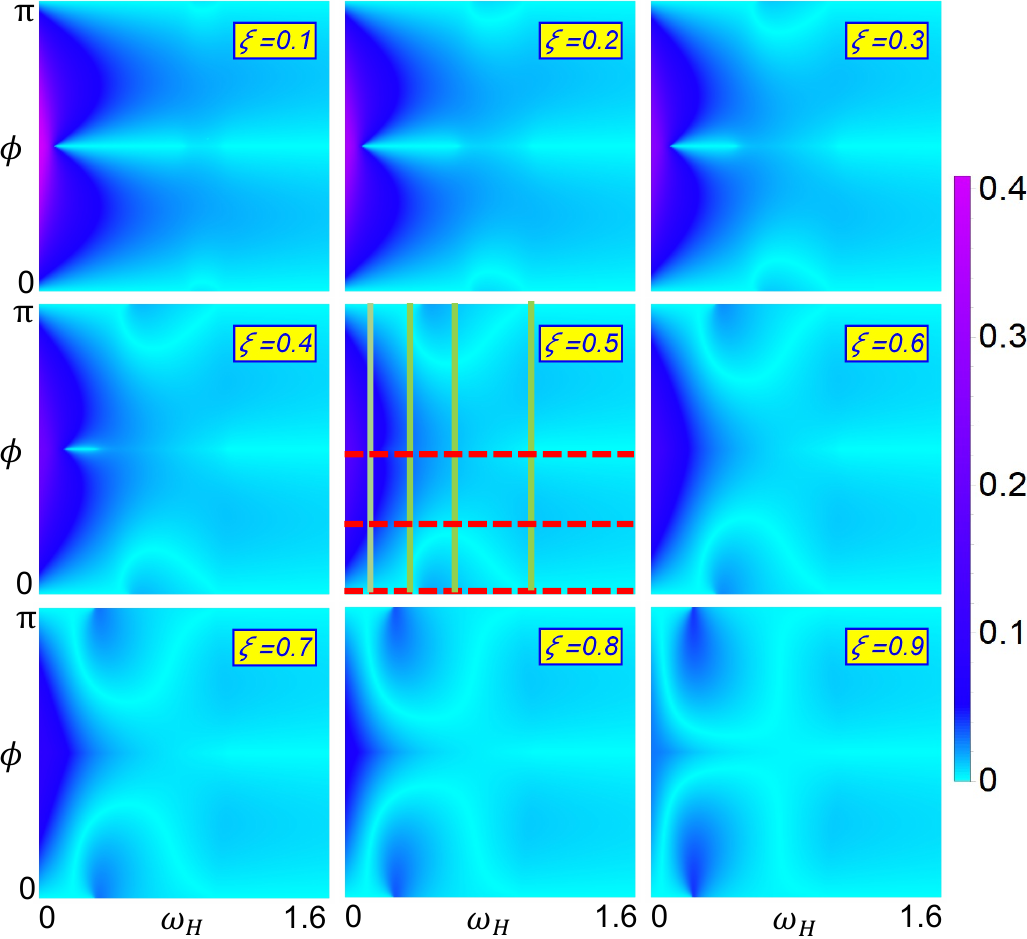}
    \caption{Density plot of $|m_{\perp}|$ as a function of $\omega_H$ (relative to $\omega_J$) and $\phi$ for nine different values of $\xi$. For $\xi=0.5$, the red and the green cuts correspond to the curves plotted in Fig.~\ref{fig:angles}(b) and Figs.~\ref{fig:angles}(c)-\ref{fig:angles}(f), respectively. The SF phase is enclosed by an ear-shaped contour near $\phi=0$ and $\pi$, which expands (shrinks) with an increasing (decreasing) $\xi$ towards the AFM (FM) limit.}
    \label{fig:xi depen}
\end{figure}

To further understand the equilibrium properties, we focus on the perpendicular component $\bm{m}_{\perp}$ where the SF phase is more conspicuous. Figure~\ref{fig:xi depen} extends Fig.~\ref{fig:angles}(e) to arbitrary field strength $\omega_H$ at different values of $\xi$, where the horizontal and vertical cuts in the special case of $\xi=0.5$ correspond to the curves plotted in Fig.~\ref{fig:angles}(b) and Figs.~\ref{fig:angles}(c)-\ref{fig:angles}(f). We can clearly see that the SF phase is enclosed by an ear-shaped contour near $\phi=0$ and $\pi$, which grows (shrinks) with an increasing (decreasing) $\xi$. For $\xi=0.1$, the SF phase is almost invisible. If the FM limit $\xi\rightarrow0$ (not shown) is reached, the SF phase will vanish identically. The variation of $\xi$ shown in Fig.~\ref{fig:xi depen} intuitively demonstrates how FIM is intrinsically connected to its FM and AFM limits at equilibrium.

To close the discussion of this section, we mention that $J$, $A$, and $K$ appearing in Eq.~\eqref{eq:free_energy} may also depend on temperature, hence implicitly depending on $\xi$. As a result, $\xi$ is not sufficient to quantitatively describe a FIM. However, the temperature dependence of $\xi$ and other parameters are material specific and quite often unclear. Our goal is to unravel the unique but universal properties of FIMs directly related to their nonequivalent sublattices rather than a case study of a specific material. Therefore, we restrict our discussion to the simplified model consisting of a single tuning parameter $\xi$.

\section{Dynamical Modes}\label{sec:modes}

\begin{figure}
    \centering
    \includegraphics[width = \linewidth]{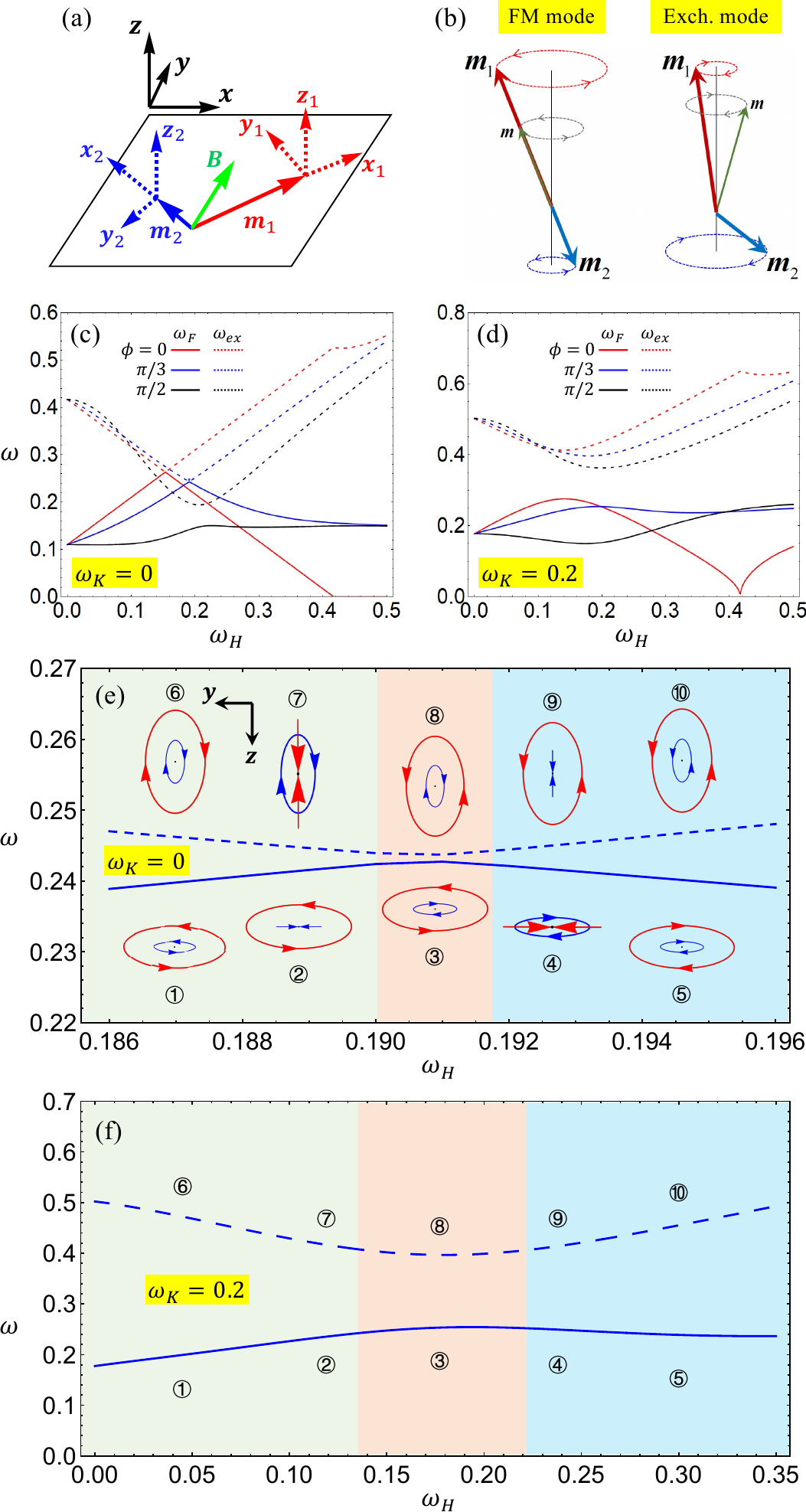}
    \caption{(a) Schematics of the model FIM in the presence of an in-plane magnetic field, where two local coordinates are defined based on the equilibrium orientations of $\bm{m}_1$ and $\bm{m}_2$. (b) Illustration of the two circularly-polarized eigenmodes in an easy-axis FIM ($\omega_K=0$) when $\bm{H}_0$ is applied along $\hat{x}$. Panels(c) and (d) plot the two eigenfrequencies as functions of the applied field along three different directions for easy-axis and easy-plane FIM, respectively, for $\xi=0.5$ (or $\beta=1/3$) and $\omega_A=0.08$. The FM mode (solid curves) and the exchange mode (dashed curves) become degenerate at a critical field below the SF threshold if and only if $\omega_K=0$ and $\phi=0$. Either a finite hard-axis anisotropy $\omega_K$ or a nonzero field angle $\phi$ (or both) will lift the degeneracy. Panels (e) and (f) are zoom-in plots of the avoided crossing corresponding to (c) and (d) for $\phi=\pi/3$, respectively. Here, the chirality of $\bm{m}_1$ ($\bm{m}_2$) is illustrated by red (blue) ellipses as seen from the $+x_1$ ($-x_2$) direction. In the FM mode, $\bm{m}_1$ ($\bm{m}_2$) becomes linearly polarized when the magnetic field reaches point \ding{175} (\ding{173}), across which $\bm{m}_1$ ($\bm{m}_2$) flips its chirality of precession. The \ding{175} and \ding{173} points separate regions of distinct elliptical precessions colored differently. The exchange mode follows a somewhat reversed pattern, which is marked by \ding{177} to \ding{181}.}
    \label{fig:modes}
\end{figure}

Having obtained the equilibrium spin configuration of the model FIM, we now turn to the dynamical eigenmodes-the way magnetic moments precess around their equilibrium positions. Even though finding the ground state can be simplified into a two-dimensional problem when $\bm{H}_0$ is confined in the easy plane, dynamical properties are intrinsically three dimensional as spin precessions unavoidably involve out-of-plane motions. In the presence of a magnetic field, $\bm{m}_1$ and $\bm{m}_2$ can become non-collinear. So, we introduce two local coordinate frames to describe their dynamical precessions around their individual equilibrium positions~\cite{Wangsness1953,Moriya1960}, which is schematically illustrated in Fig.~\ref{fig:modes}(a). Under this geometry, the free energy in Eq.~\eqref{Ham} can be rewritten as
\begin{align}
 \mathcal{E}& = \frac{\hbar\omega_J \xi}{(\xi+1)^2} \left[Z_1Z_2 + (X_1X_2 + Y_1Y_2)\cos{(\theta_1-\theta_2)} \right. \notag\\
 & \left.+ (X_1Y_2 - Y_1 X_2)\sin{(\theta_1-\theta_2)}\right] + \frac{\hbar\omega_K}{2(\xi+1)^2}(Z_1^2+\xi^2Z_2^2) \notag\\
 & -\frac{\hbar\omega_A}{2(\xi+1)^2}\left[X_1\cos^2\theta_1+Y_1\sin^2\theta_1 \right. \notag\\
 & \qquad+\xi^2(X_2\cos^2\theta_2+Y_2\sin^2\theta_2^2) \notag\\
 & \qquad\left. -2(X_1Y_1\cos\theta_1\sin\theta_1+\xi^2X_2Y_2\cos\theta_2\sin\theta_2)\right] \notag\\
 & - \frac{\hbar\omega_H}{\xi+1}\left[X_1^2\cos(\phi - \theta_1 )+\xi X_2^2\cos (\phi - \theta_2) \right. \notag\\
 & \qquad \left. -Y_1^2\sin(\phi - \theta_1 )+\xi Y_2^2 \sin(\phi - \theta_2 )\right],
\end{align}
where $X_i$, $Y_i$, and $Z_i$ stand for the components of $\bm{m}_i$ ($i=1, 2$) normalized in the associated local coordinate frame. Writing the Landau-Lifshitz-Gilbert (LLG) equation $\dot{\bm{m}}_i=\bm{f}_i\times\bm{m}_i$ with $\bm{f}_i=-\delta\mathcal{E}/\hbar\delta\bm{m}_i$ in terms of $X_i$, $Y_i$, and $Z_i$, we are able to linearize the spin dynamics to obtain the eigenfrequencies and the eigenmodes. Including the Gilbert damping $\alpha\bm{m}_i\times\dot{\bm{m}}_i$ causes a slight frequency shift while the essential feature of the spectrum is kept, so in this section we omit the damping effect for simplicity. However, damping effects will become crucial in the next section.

We begin with the simplest case where the hard axis anisotropy vanishes ($\omega_K=0$) and $\bm{H}_0$ is applied along the easy axis ($\phi=0$). In this circumstance, the system assumes rotational symmetry around the easy axis, which guarantees a collinear ground state. Similar to the collinear AFM case, the rotational symmetry leads to two circularly polarized modes. Their eigenfrequencies are solved as (in the exchange approximation $\omega_A\ll\omega_J$) 
\begin{subequations}
\begin{align}
 &\omega_{F} = \frac{\sqrt{\omega_{A}^{2} + 2 \omega_J \omega_A + \beta^2 \omega_{J}^{2}}}{2} -\frac{\beta}{2} \left( \omega_J - \omega_A \right) + \omega_H, \\
 &\omega_{ex}= \frac{\sqrt{\omega_{A}^{2} + 2 \omega_J \omega_A + \beta^2 \omega_{J}^{2}}}{2} +\frac{\beta}{2} \left( \omega_J - \omega_A \right) - \omega_H,
\end{align}
\label{eq:eigenfreq}%
\end{subequations}
where the subscript ``F" (``ex") indicates the FM (exchange) mode lying in the GHz (sub-THz) regime, and $\beta \equiv \frac{1-\xi}{1+\xi}$ ranges from $0$ (AFM limit) and $1$ (FM limit). In the AFM limit $\beta\rightarrow0$ (or $\xi\rightarrow1$), Eq.~\eqref{eq:eigenfreq} reduces to $\omega_{\pm}=\sqrt{\omega_A^2+2\omega_J\omega_A}/2\pm\omega_H$, reproducing Kittel's formula for AFM materials. Here, the additional factor of $1/2$ originates from the definition in Eq.~\eqref{eq:spinratio}: $|M_i|/M_s\rightarrow1/2$ as $\xi\rightarrow1$. In the FM limit $\beta\rightarrow1$ (or $\xi\rightarrow0$), we have $\omega_F=\omega_H$ while $\omega_{ex}=\omega_J-\omega_H$ if $\omega_A\rightarrow0$. Figure~\ref{fig:modes}(b) schematically illustrates these two chiral modes for $\beta\rightarrow1$ and $\omega_A\rightarrow0$ as follows. (1) The FM mode exhibits a right-handed precession in which the two magnetic moments are kept collinear. Even though a finite $\omega_A$ will make the cone angle of $\bm{m}_1$ (the longer spin) slightly larger than that of $\bm{m}_2$ (the shorter spin), this mode is essentially driven by the Zeeman interaction, hence a lower frequency. (2) By contrast, the exchange mode exhibits a left-handed precession in which $\bm{m}_2$ has an apparent larger cone angle than $\bm{m}_1$, leveraging the strong exchange interaction to drive the precession of the magnetic moments, hence a much higher frequency. In general, when neither limit is close, the two modes cross each other at a field below the SF transition, as plotted by the red curves in Fig.~\ref{fig:modes}(c).

If the magnetic field is tilted away from the easy axis, the rotational symmetry will be broken. As a result, the two circularly polarized modes will hybrid as their eigenfrequencies approach, leading to an avoided crossing as shown by the blue and black curves in Fig.~\ref{fig:modes}(c). This anti-crossing gap can be enlarged appreciably by a finite hard-axis anisotropy $\omega_k$, which also breaks the rotational symmetry, as plotted in Fig.~\ref{fig:modes}(d).

To better understand the hybridization of the chiral eigenmodes in the absence of rotational symmetry, we zoom in on the vicinity of the anti-crossing gap and show the evolution of chirality for each magnetic moment in their local frame in Figs.~\ref{fig:modes}(e) and~\ref{fig:modes}(f). For the FM mode, $\bm{m}_1$ ($\bm{m}_2$) flips its chirality as the magnetic field crosses point \ding{175} (\ding{173}), which separates regions of opposite chirality for $\bm{m}_1$ ($\bm{m}_2$). The exchange mode follows a somewhat reversed pattern, where $\bm{m}_1$ flips its chirality first, followed by $\bm{m}_2$, with a sweeping magnetic field. A hard axis anisotropy enlarges the gap and broadens the window in which the two magnetic moments exhibit opposite chirality, i.e. \ding{174} for the FM mode and \ding{179} for the exchange mode.

\begin{figure}[t]
    \centering
    \includegraphics[width = \linewidth]{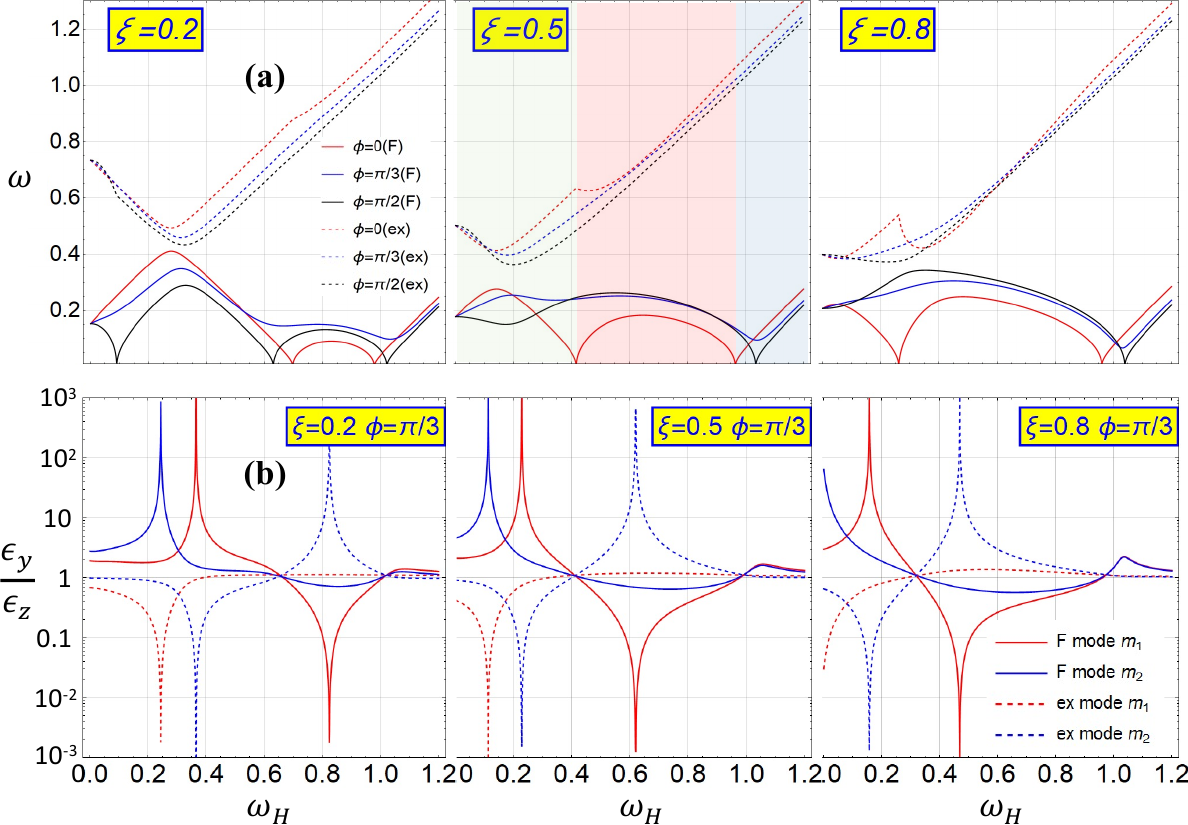}
    \caption{(a) Eigenfrequencies versus magnetic field along different in-plane directions for $\xi=0.2$, $0.5$, and $0.8$. The collinear, SF, and spin-flip phases are shaded in different colors for the case of $\xi=0.5$ and $\phi=0$. 
    (b) Polarization of each sublattice magnetic moment expressed as the ratio of the principal axes of elliptical trajectory in the local frame for $\phi=\pi/3$ at corresponding values of $\xi$. The diverging (vanishing) locations indicate linear polarization along the local in-plane $Y$ (out-of-plane $Z$) direction. The first two divergences correspond to the two chirality flips depicted in Fig.~\ref{fig:modes}. The third one takes place inside the SF phase.}
    \label{Fig: whole phases}
\end{figure}

Finally, we study the influence of $\xi$---the central parameter in our model---on the eigenmodes. Figure~\ref{Fig: whole phases}(a) plots the two eigenfrequencies as functions of the sweeping magnetic field along different in-plane directions for $\xi=0.2$, $0.5$, and $0.8$, where for the case of $\xi=0.5$ we have colored the collinear, SF, and spin-flip phases differently. In Fig.~\ref{Fig: whole phases}(b), we plot the polarization of each sublattice magnetic moment in terms of the ratio of principal axes of the elliptical trajectory in the local frame, $\epsilon_y/\epsilon_z$, for $\phi=\pi/3$ at the three corresponding values of $\xi$. The diverging (vanishing) locations are where the magnetic moment becomes linearly polarized along the in-plane $Y$ (out-of-plane $Z$) direction in the local frame. Since now the plot has been extended to a much higher field compared with Fig.~\ref{fig:modes}, an additional chirality flip taking place in the SF phase shows up.

%-------------------------------------------------------------------------------------------------------------------------------------------------%

\section{Spin-Torque Oscillators}\label{sec:STO}

Given the static properties and dynamical modes, it is natural to ask how FIMs react to, thus being manipulated by, applied currents. Similar to those in FM and AFM materials, the chiral dynamical modes can be excited by absorbing angular momenta from external currents through dampinglike torques. For instance, if a dampinglike torque competes with the intrinsic Gilbert damping (anti-damping effect), it will pump energy and angular momenta against equilibrium into the system. When its strength surpasses a threshold, the Gilbert damping will be overwhelmed, which then triggers a dynamical instability that typically manifests as a spontaneous oscillation (also known as auto-oscillation) of the magnetic moments. By realizing this phenomenon, the current-controlled magnetic system is regarded as a spin-torque(ST) oscillator, which converts dc inputs into high-frequency ac outputs. While ST oscillators have been widely studied in FM and AFM materials~\cite{tsoi1998excitation,tsoi2000generation,KIM2012217,AkermanReview,demidov2012magnetic,cheng2016terahertz,khymyn2017antiferromagnetic}, they are much less understood in FIM materials, especially when the ratio of sublattice moments can be varied continuously. While previous theoretical investigations of ST oscillators in FIM materials are conducted near the compensation point~\cite{lisenkov2019subterahertz,cutugno2021micromagnetic}, we explore the behavior of ST oscillators in a wide range of ratio of sublattice spins.

By navigating $\xi$ freely between the FM and AFM limits, we strive to understand the ST-induced oscillations in FIMs as an interposition smoothly connected to its FM and AFM counterparts. In particular, we will figure out whether exciting the sub-THz exchange mode can emulate the performance of an ultrafast AFM system. To this end, we focus on the dampinglike ST which can be generated by either the spin Hall effect or the interfacial Rashba spin-orbit coupling. Specifically, the ST acting on the sublattice moment $\bm{m}_i$ ($i=1,2$) is $\bm{\tau}_i = \bm{m}_{i} \times (\bm{\omega}_{s} \times \bm{m}_{i})$~\cite{lisenkov2019subterahertz,woo2018current}, where $\bm{\omega}_{s}=\omega_s\bm{p}$ and $\omega_s$ is proportional to the current-induced spin accumulation scaled into the frequency dimension. Here $\bm{p}$ is the unit vector of magnetic moment associated with the current-induced spin accumulation which adsorbs the $\gamma$ factor. The spin dynamics is then described by the coupled LLG equation
\begin{equation}\label{eq:LLG}
\dot{\bm{m}}_i = \bm{f}_{i} \times \bm{m}_{i} + \frac{\alpha}{m_{i}} \bm{m}_{i} \times \dot{\bm{m}_{i}} + \bm{\tau}_i,
\end{equation}
where $\bm{f}_i=-\delta\mathcal{E}/\hbar\delta\bm{m}_i$ and $\alpha$ is the Gilbert damping constant. The damping term comes with a ratio $\alpha/m_{i}$ to ensure the conservation of angular momentum. In the following, we focus on field-free dynamics, so $\omega_{H} = 0$. Other parameters are given representative values $\omega_{A} = 10^{-4}\omega_{J}$ and $\alpha = 0.005$. (Note that in the previous section, $\omega_{A}$ has been amplified for demonstration purposes.) Since $x$ is the easy axis, we use the initial conditions $\bm{m}_{1}\parallel +x$ and $\bm{m}_{2}\parallel -x$. Since the free energy density Eq.~\ref{eq:free_energy} uses macrospin approximation, we do not consider non-uniform spin dynamics.

\begin{figure}
    \centering
    \includegraphics[width = \linewidth]{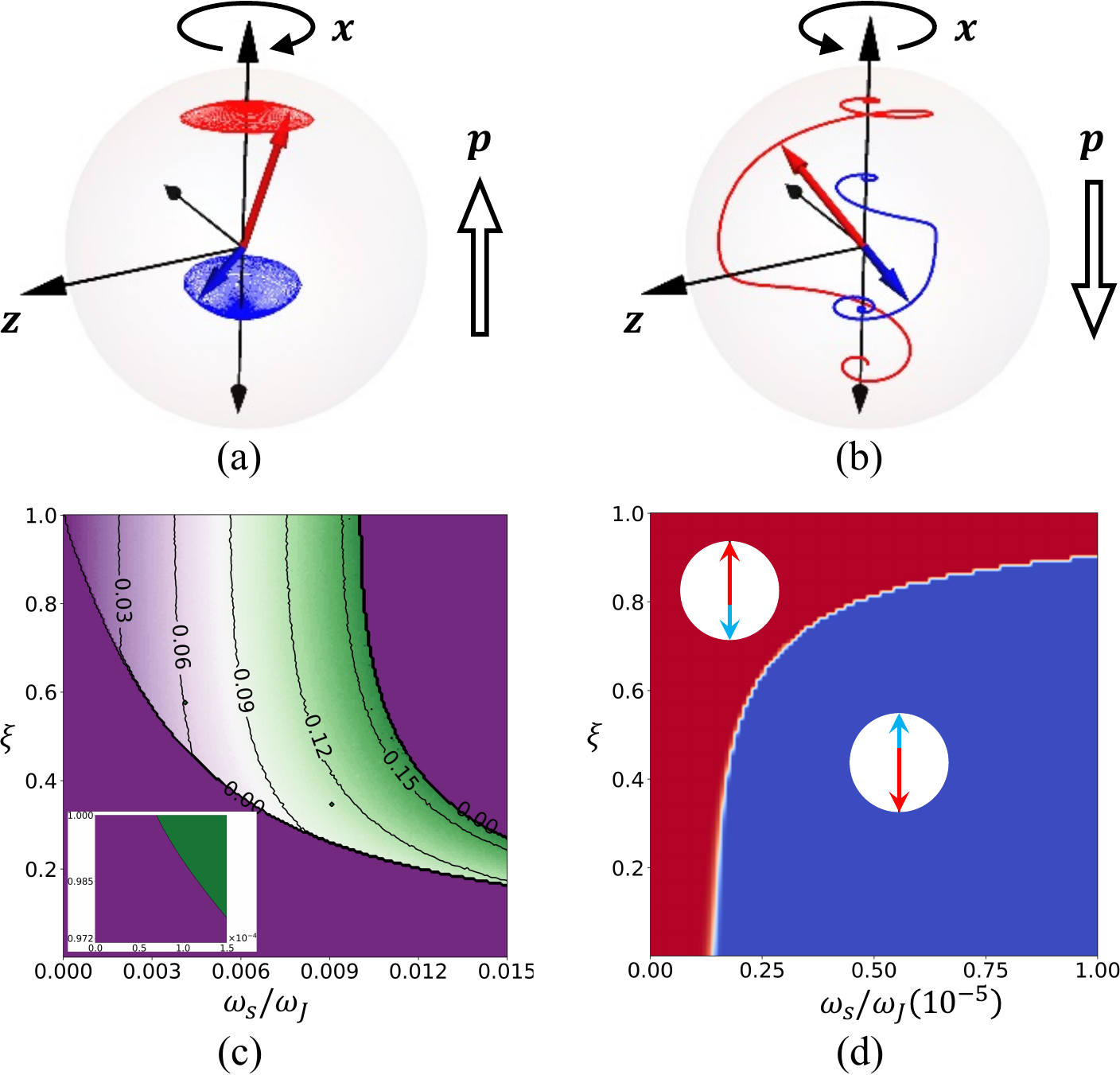}
    \caption{ST-driven dynamics of chiral modes. Trajectories of magnetic moments for $\bm{p}\parallel x$ (a) and $\bm{p}\parallel -x$ (b), where the left-handed exchange mode grows and saturates at a steady-state oscillation whereas the right-handed FM mode evolves into a magnetic switching. (c) Terminal frequency of the exchange mode as a function of $\omega_s$ and $\xi$, where the threshold diverges as the FM limit is approached. Inset: zoom-in display of the phase boundary near $\xi\rightarrow1$. (d) Terminal spin configuration as a function of $\omega_s$ and $\xi$, where the switching threshold diverges in the AFM limit $\xi\rightarrow1$. The easy-axis anisotropy is taken to be $\omega_A=10^{-4}\omega_J$.}
    \label{fig:STO_nohard_x_pol}
\end{figure}

We first revisit the two chiral modes discussed in the preceding section and figure out how they react to the ST. To this end, we set $\omega_K=0$ and consider $\bm{p}$ along the $\pm x$ axis. Figures~\ref{fig:STO_nohard_x_pol}(a) and~\ref{fig:STO_nohard_x_pol}(b) show the simulated trajectories of an FIM with $\xi=0.5$ driven by the ST with $\bm{p}\parallel x$ and $\bm{p}\parallel-x$, respectively. In the former, the ST excites the left-handed exchange mode whose amplitude grows towards a saturation, arriving at a steady-state oscillation of sub-THz frequency. In the latter, the right-handed FM mode is initiated, which grows without bound and inevitably evolves towards a magnetic switching. Figures~\ref{fig:STO_nohard_x_pol}(c) and~\ref{fig:STO_nohard_x_pol}(d) are the dynamical phase diagrams corresponding to~\ref{fig:STO_nohard_x_pol}(a) and~\ref{fig:STO_nohard_x_pol}(b), respectively. In Fig.~\ref{fig:STO_nohard_x_pol}(c), the terminal frequency of auto-oscillation is plotted as a function of the applied ST $\omega_s$ and $\xi$, where in the AFM limit ($\xi\rightarrow1$) the threshold ST reduces to $\omega_s^{\rm th}=\alpha\sqrt{\omega_A\omega_J}$~\cite{Cheng2014} (see the enlarged plot in the inset) and in the FM limit ($\xi\rightarrow0$) the auto-oscillation threshold disappears completely. This feature is quite understandable because the exchange mode does not exist in the FM limit; it is intrinsically a multisublattice property. If the driving ST $\omega_s$ is too large, however, the system will undergo an ST-induced SF transition without a sustainable auto-oscillation~\cite{cheng2016terahertz,GomonayThreshold}. In Fig.~\ref{fig:STO_nohard_x_pol}(d), the terminal states are separated into the switching (blue) and nonswitching (red) regions, where the switching threshold diverges for $\xi\rightarrow 1$, i.e. no switching in the AFM limit. In fact, the threshold cannot diverge because a sufficiently large $\omega_s$ will lead to an SF transition.

\begin{figure}[t]
    \centering
    \includegraphics[width = \linewidth]{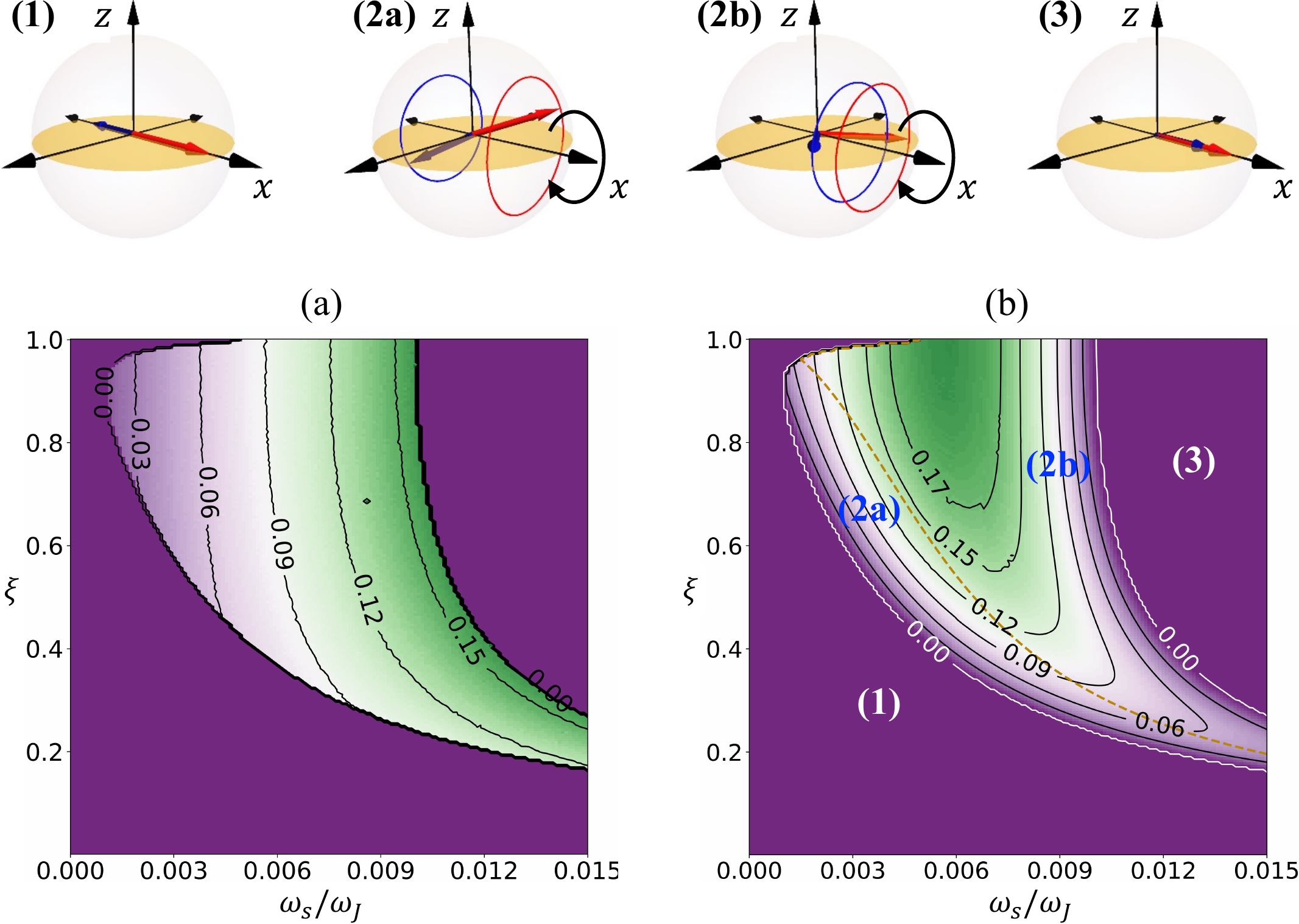}
    \caption{(a) Output frequency (normalized to $\omega_J$) and (b) DC spin pumping (in arbitrary unit) of an ST oscillator in the presence of hard-axis anisotropy (along $z$) driven by an in-plane spin accumulation (along $x$) as functions of $\omega_{s}$ and $\xi$. Panels (1)-(3) are schematics of the spin configurations and trajectories associated with different phases in (b), where sub-phases (2a) and (2b) are separated by the dotted orange curve.}
    \label{fig:STO_x_pol}
\end{figure}

Next, we turn on the hard axis anisotropy and take $\omega_{K} = 10^{-2} \omega_{J}$ as a representative value to study the dynamical behavior of the high-frequency left-handed mode. Besides the auto-oscillation frequency shown in Fig.~\ref{fig:STO_x_pol}(a), we also plot the dc component of spin pumping $I_{dc}=g_{\uparrow\downarrow}\hat{\bm{x}}\cdot\sum_i\bm{m}_i\times\dot{\bm{m}}_i$ in Fig.~\ref{fig:STO_x_pol}(b), where we assumed that the spin-mixing conductance $g_{\uparrow\downarrow}$ is the same for both sublattices. The legitimacy of the form of $I_{dc}$ will be discussed in the following section. Comparing Fig.~\ref{fig:STO_nohard_x_pol}(c) with Fig.~\ref{fig:STO_x_pol}(a), we see that the introduction of hard-axis anisotropy considerably changes the phase boundary in the AFM limit $\xi\rightarrow1$. The output frequency alone is insufficient to capture the physics of the auto-oscillation; we also need to look into the topology of trajectories. In this regard, we identify two subphases of auto-oscillation interpolating the static phase (1) and the SF phase (3), which are schematically illustrated by (2a) and (2b) as labeled in Fig.~\ref{fig:STO_x_pol}(b). While both sublattice magnetic moments precess with the left-handed chirality, their trajectories lie on the same (different) hemisphere(s) in phase (2b) [phase (2a)]. With increasing ST strength, the output frequency increases monotonically as the system evolves from phase (2a) to (2b), whereas the dc spin pumping is apparently non-monotonic and the maximum value takes place inside phase (2b). This complicated behavior is attributed to the fact that spin pumping not only depends on the projection of magnetic moments onto the quantization axis but is also proportional to the output frequency.

Compared with the case of uniaxial FIM shown in Fig.~\ref{fig:STO_nohard_x_pol}(a) and~\ref{fig:STO_nohard_x_pol}(c), the auto-oscillation threshold [phase boundary separating (1) and (2a)] in the presence of hard-axis anisotropy reaches a minimum at about $\xi=0.87$---a value depending on the strength of the hard-axis anisotropy $\omega_K$. In the AFM limit ($\xi\rightarrow1$), phase (2a) disappears so that the oscillator jumps directly from the static phase (1) to phase (2b), which has been discussed for AFM nano-oscillators~\cite{cheng2016terahertz}. A larger $\omega_K$ results in a larger reduction of the threshold at the minimum compared to what it is in the AFM limit. When $\omega_K$ becomes even larger, phase (2b) on the AFM line $\xi=1$ can be destroyed so that the oscillator undergoes an ST-induced SF transition from (1) to (3), skipping any auto-oscillation phase. This is because we have not included any feedback mechanism which is essential to stabilize an ST oscillator by allowing its amplitude to be controlled continuously by $\omega_s$. Away from the AFM limit, however, an auto-oscillation phase is always achievable regardless of $\omega_s$ or feedback mechanism. In other words, an FIM oscillator can stabilize on its own and afford an ST-driven auto-oscillation without external feedback. The lowered threshold ST plus the self-sustainability are unique advantages of FIMs, which can function as a high-frequency AFM when the exchange mode is stimulated. It should be noted that in our work both easy-axis and hard-axis anisotropy are included, while previous works only considered either easy-axis or hard-axis anisotropy but not both~\cite{lisenkov2019subterahertz,cutugno2021micromagnetic}, and, more importantly, $\xi$ does not span the full range between the FM and AFM limits.

To further understand the dynamical behavior of FIM oscillators, we now derive the threshold ST by linearizing the LLG equations~\eqref{eq:LLG}, which yields the characteristic equation
\begin{align}\label{eq:STO_eigen_freq}
(\omega_{I}^{2} + &\alpha \omega_J \omega_{I} + f_{A}) (\omega_{I}^{2} + \alpha \omega_J \omega_{I} + f_{K})  \notag \\
&+ \left( \frac{\omega_J\omega_s}{2} + \Delta \omega_{J} \omega_{I} \right)^{2} = 0,
\end{align}
where $\omega_{I} = i \omega + \Delta \omega_s/2$ with $\omega$ the (complex valued) eigenfrequency and $\Delta\omega_s=\omega_s|m_1-m_2|$, and $2f_{A (K)} = -\alpha \omega_J \Delta \omega_{s}  + \omega_J \omega_{A(K)} ( 1 + |m_1-m_2|^2 ) $. While the solution to Eq.~\eqref{eq:STO_eigen_freq} in the AFM limit reduces to $2\omega=i\alpha\omega_J\pm[(\omega_A+\omega_{K})\omega_J-\alpha^2\omega_J^2\pm\omega_J\sqrt{(\omega_K-\omega_A)^2-4\omega_s^2}]^{1/2}$, which is consistent with the AFM result~\cite{cheng2016terahertz,GomonayThreshold}, the general solution is too complicated in expression unless we set $\omega_K=0$. The imaginary part $\rm{Im}(\omega)$ determines the dissipation of the oscillator, which in general results from the competition between the Gilbert damping and the dampinglike ST. When $\omega_{s}$ exceeds a critical value, $\rm{Im}(\omega)$ will flip sign, marking the onset of auto-oscillation. Therefore, solving $\rm{Im}(\omega)=0$ gives the phase boundary of auto-oscillation in Fig.~\ref{fig:STO_x_pol}. 

\begin{figure}[t]
    \centering
    \includegraphics[width = \linewidth]{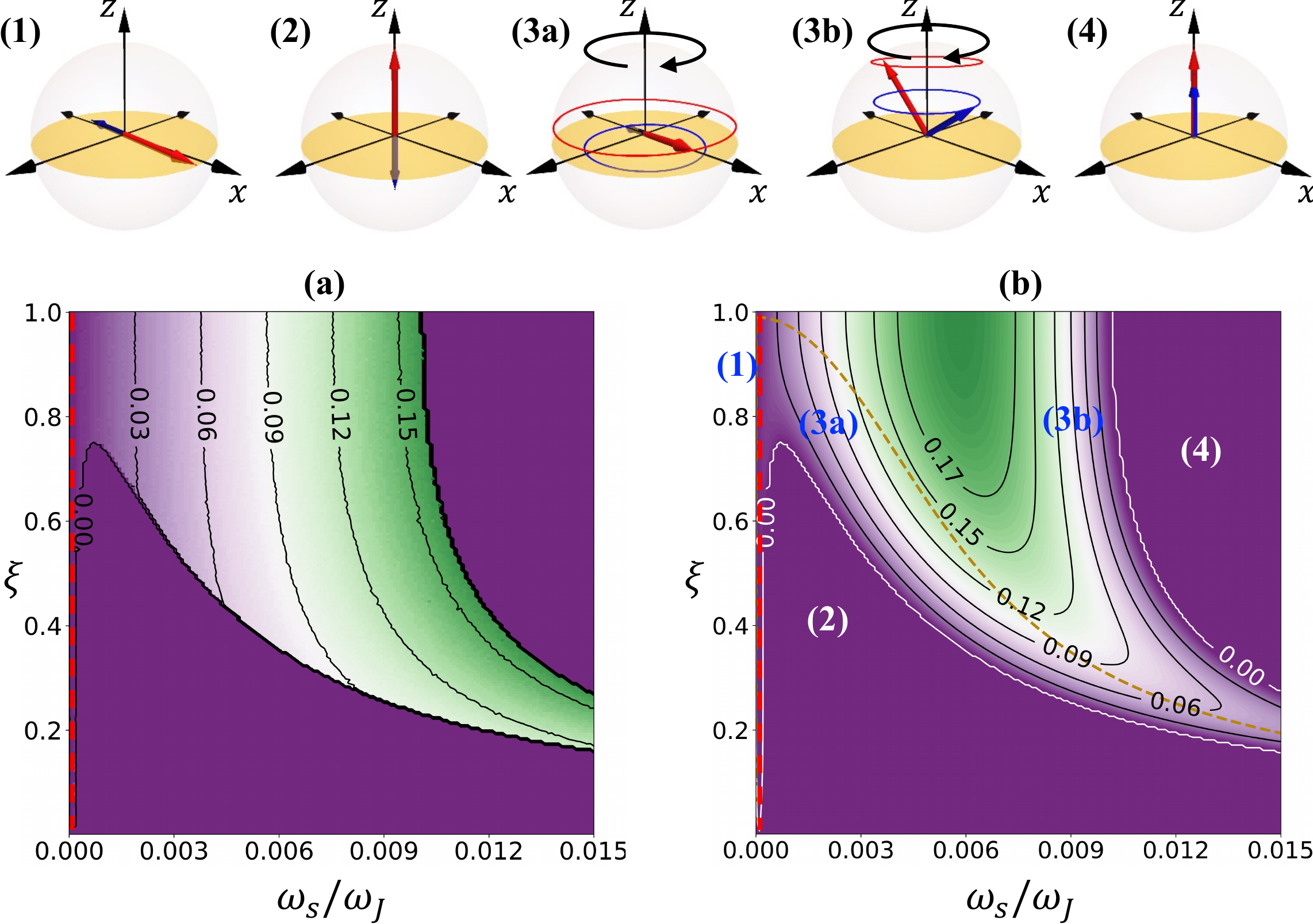}
    \caption{(a) Output frequency and (b) dc spin pumping of an ST oscillator driven by an out-of-plane spin accumulation as functions of $\xi$ and $\omega_{s}$. The red dashed lines marking the static phase (1) are added for visual clarity (because $\omega_A=10^{-4}\omega_J$ is very small, so is the threshold). Panels (1)--(4) are schematics of the spin configurations and trajectories associated with different regions in (b), where subphases (3a) and (3b) are separated by the dotted orange curve.}
    \label{fig:STO_z_pol}
\end{figure}

Finally, we consider the case of out-of-plane spin accumulation (i.e. $\bm{p}$ along $+z$ axis), which can be generated by polarizing the electrons with a perpendicularly magnetized FM layer in conjunction with the FIM layer. But again, different from existing studies~\cite{lisenkov2019subterahertz}, we considered the full range of $\xi$ connecting the FM and the AFM limit rather than the neighborhood of the compensation point. The numerical result is shown in Fig.~\ref{fig:STO_z_pol} using the same parameters as those in Fig.~\ref{fig:STO_x_pol}. Compared to the case of in-plane spin accumulation, the present situation exhibits more dynamical phases. In Fig.~\ref{fig:STO_z_pol}, phase (1) is the static phase where the magnetic moments do not move but are slightly perturbed away from the easy axis. Since the in-plane anisotropy in the simulation is very small ($\omega_A=10^{-4}\omega_J$), it is hard to visualize this static phase, so we manually add red dashed lines in Figs.~\ref{fig:STO_z_pol}(a) and~\ref{fig:STO_z_pol}(b). With larger $\omega_{s}$, the magnetic moments are able to precess on different hemispheres and the system reaches phase (3a). Further increasing $\omega_{s}$, we observe complicated behavior with varying $\xi$: when $\xi$ is smaller than about 0.75, there exists a spin-reorientation phase [phase (2)] in which the two magnetic moments are switched from the easy $x$ axis to the hard $z$ axis without auto-oscillation. When $\xi$ is larger than about 0.75, there are two subphases of distinct topology of precessional trajectories in the auto-oscillation region labeled by (3a) and (3b), similar to what is shown in Fig.~\ref{fig:STO_x_pol}. However, the rotational axis in this case is the $z$ axis rather than the $x$ axis.

\section{Discussions}\label{sec:discuss}

In the preceding section, we assumed that the dampinglike torques act on different sublattice magnetic moments independently. That is to say, we did not include cross terms such as $\bm{m}_1\times(\bm{\omega}_s\times\bm{m}_2)$ intertwining the two sublattices. While this assumption is valid in the AFM limit as it satisfies the combined symmetry under spin-flip and sublattice exchange operations, it becomes questionable when $m_1\neq m_2$. Correspondingly, the form of spin pumping will involve cross terms like $\bm{m}_1\times\dot{\bm{m}}_2$~\cite{Kamra2017PRL,Brataas2018PRB}. When the intersublattice contributions to the ST and spin pumping are considered, the result shown in Fig.~\ref{fig:STO_x_pol} can be very different. However, it remains an open question as to how strong the intersublattice terms may become in a specific system, which calls for a detailed calculation of the spin-dependent scattering on the interface. The central point of this paper is that an FIM with variable ratio of sublattice moments can exhibit highly nontrivial dynamics even with the simplest form of ST. Moreover, we did not include the fieldlike torques in our study. The fieldlike torques do not compete with the Gilbert damping and thus do not induce auto-oscillations. But they can shift the eigenfrequency of an oscillator similar to an applied magnetic field. The relative strength of dampinglike and fieldlike torques depend on how the ST is created. If the ST arises from the spin Hall effect in a heavy metal, the fieldlike component is indeed negligible, which is supported by recent experiments~\cite{seung2017temperature}.

In real FIM materials such as the alloys of Fe, Gd, and Co, the gyromagnetic ratios for the two sublattices are typically $10\%$ different~\cite{KittelReview,ScottReview}, which may lead to deviations from our predictions based on equal gyromagnetic ratios. Nevertheless, concerning spin pumping and ST-driven dynamics discussed in the previous section, our theory remains essentially valid because the interfacial spin transmissions respect the conservation of angular momenta rather than that of the magnetic moments. Therefore, if we interpret $\xi$ in the dynamical phase diagrams shown in the previous section as the ratio of spins rather than the magnetic moments, our predictions are still applicable even though the phase boundary may be subject to a minor deformation. What could be slightly different for the case of $\gamma_1\neq\gamma_2$ is that the net magnetization does not fully compensate even when $\xi\rightarrow1$.

Finally, the present study can be generalized into noncollinear ferrimagnets with more than two sublattices. We notice that coherent spin dynamics in noncollinear AFM materials aroused attention recently, where spin pumping~\cite{lund2021spin} and spin-torque oscillators~\cite{shukla2021spintorquedriven} had been studied at a similar level of depth as their counterparts in collinear AFM materials. On the same basis, we anticipate that noncollinear FIMs will be an active direction of research in the near future. Furthermore, going beyond uniform spin dynamics by considering the spin-wave excitations (for finite momenta $\bm{k}$) will be an important subject. In particular, the role of varying $\xi$ for spin waves will be a key issue.

\begin{acknowledgments}
This work is supported by the Air Force Office of Scientific Research under Grant No.FA9550-19-1-0307. R.C. acknowledges fruitful discussions with K. Belashchenko, M. Kläui, M. Wu, and Q. Shao.
\end{acknowledgments}

\bibliography{citation}

\end{document}